\documentclass[a4paper,11pt,final]{article}

\usepackage{epsfig}
\usepackage[cmex10]{amsmath}
\usepackage{amssymb,amsthm}
\usepackage{mathtools}
\usepackage{nccmath}
\usepackage{soul}
\usepackage[normalem]{ulem}
\usepackage{cite}
\usepackage{todonotes}
\usepackage{tikz}
\usepackage{hyperref}
\usetikzlibrary{hobby,decorations.markings,arrows,intersections,shapes}
\usepackage{pgfplots}
\usepgfplotslibrary{fillbetween}
\pgfplotsset{compat=newest}
\usepgfplotslibrary{fillbetween}
\usepackage{verbatim}

\begin{document}

\title{\ \\ \LARGE\bf Do Random and Chaotic Sequences Really Cause Different PSO Performance?}

\author{Paul Moritz N\"orenberg and Hendrik~Richter \\
HTWK Leipzig University of Applied Sciences \\ Faculty of
Engineering \\
        Postfach 301166, D--04251 Leipzig, Germany. \\ Email:
        paul\_moritz.noerenberg@stud.htwk-leipzig.de\\ 
hendrik.richter@htwk-leipzig.de. }

\maketitle

\begin{abstract}

Our topic is  performance differences between using random and chaos for  particle swarm optimization (PSO). We take random sequences with different  probability distributions and compare them to chaotic sequences with different but also with  same density functions. This enables us to differentiate between differences in the origin of the sequences (random number generator or chaotic nonlinear system) and statistical differences expressed by the underlying distributions. Our findings (obtained by evaluating the PSO performance for various benchmark problems using statistical hypothesis testing) cast considerable doubt on previous results which compared random to chaos and suggested that the choice leads to intrinsic differences in performance.

\end{abstract}

\section{Introduction}
If and under what circumstances chaotic sequences give different performance results than random sequences in evolutionary computation is a topic of intensive debate~\cite{capo03,chen18,gag21,xu18,ma19,gan13,tian19,plu14,zel22}. This debate has mainly been conducted by taking a chaotic map and comparing for a set of benchmark problems the performance results obtained by sequences from the map  with the results involving pseudo random number generators (PRNGs). In most of these works we find the claim that chaos somehow improves the performance as compared to random but different results have also been reported~\cite{ala09,chen18,gan13,tian19,plu14,zel22,kuang14,rong10,yang12}. With the focus on particle swarm optimization (PSO), this paper adds to the discussion but proposes a different approach. 

Suppose there are differences in PSO performance between using chaotic and random sequences and given the PSO implementations and parameters are the same except the sequences used. Then, these differences in performance must correspond to differences between the sequences, and  a question worth asking is how these sequences could possibly differ. Apart from the difference in origin of both types of sequences, the main reason which could make chaotic sequences different from random ones concerns the sequences as a whole  and involves differences in the underlying distribution or density.

\begin{figure}[htb]
\begin{center}
\includegraphics[trim = 35mm 90mm 42mm 100mm,clip, width=6.2cm, height=5.2cm]{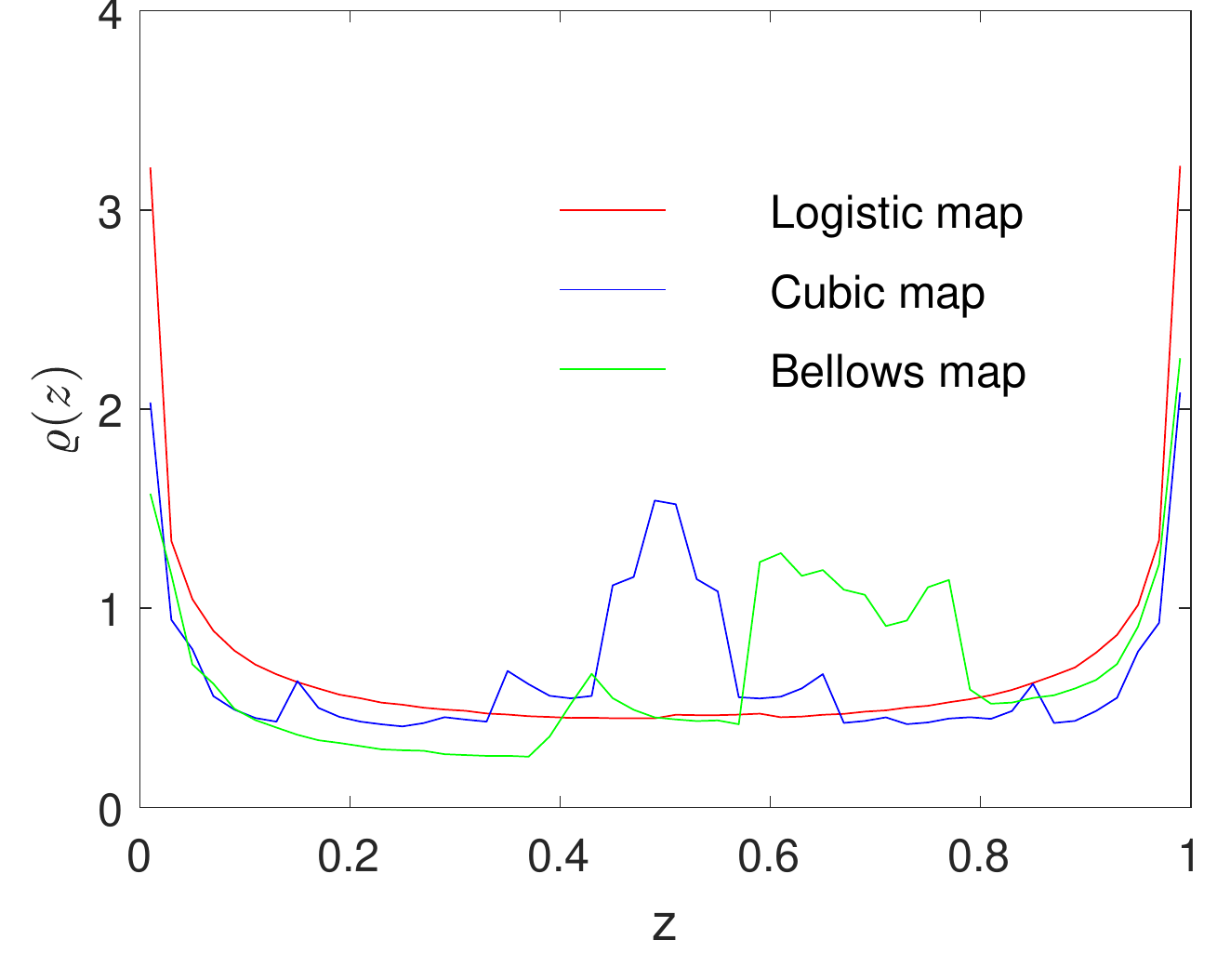}
\includegraphics[trim = 35mm 90mm 42mm 100mm,clip, width=6.2cm, height=5.2cm]{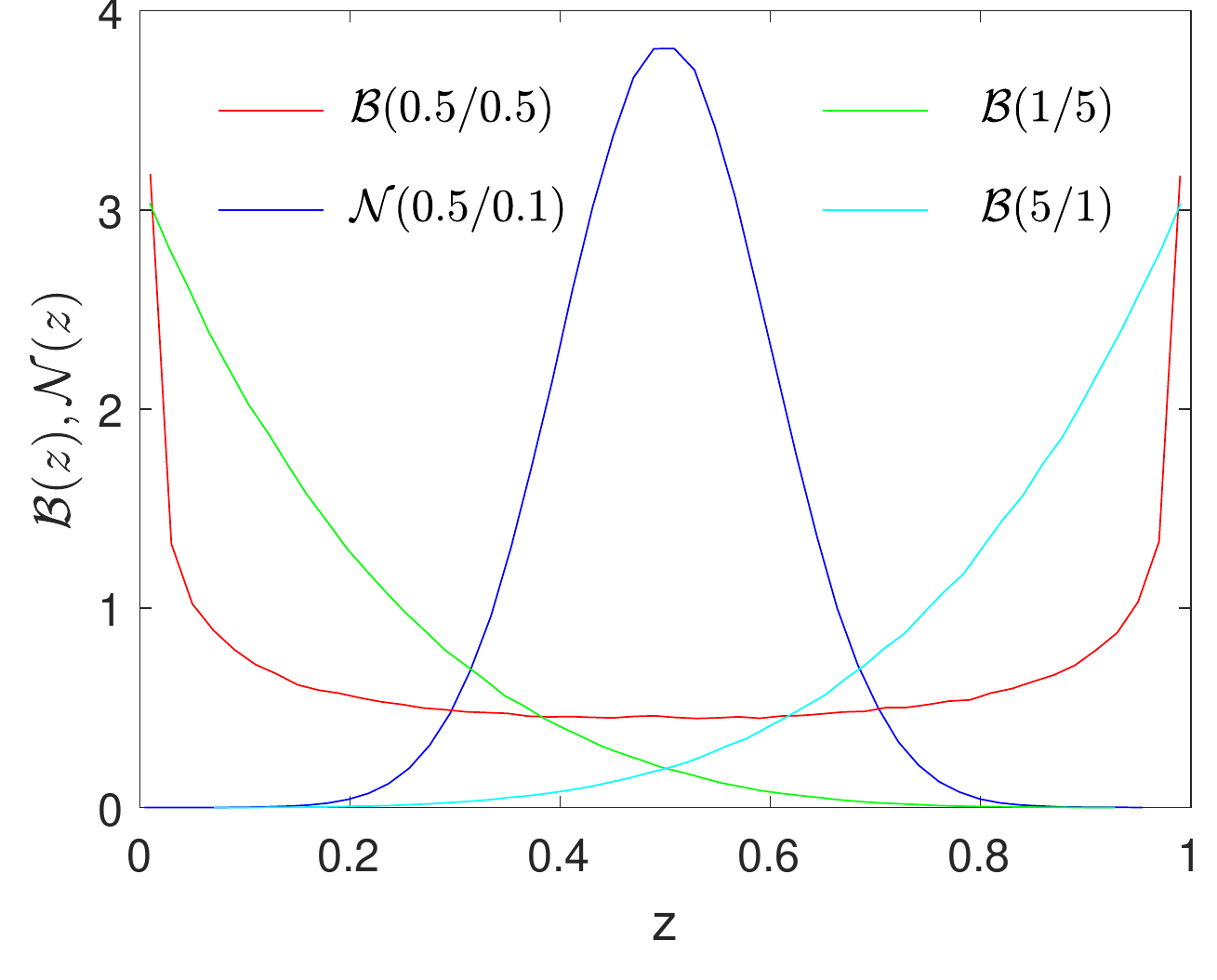}

(a) \hspace{4cm} (b)
\caption{Distribution of sequences used in the study (a) Natural invariant densities of chaotic maps. (b) Probability density functions of random variables. }
\label{fig_chaotic}
\end{center}
\end{figure}

Our addition to the discussion about performance differences between chaotic and random sequences is to 
disentangle the effect of the underlying distribution and effects connected to how the sequences are generated (PRNGs or chaotic maps).
It is well-known that PRNGs can produce
random sequences associated with different  probability distributions, thus offering to study the effect that  different distributions have on PSO performance, see Figure~\ref{fig_chaotic}(a) for the distributions considered in this study. A first observation is that
chaotic sequences obtained from a certain variety of nonlinear maps also have different  invariant densities, thus enabling to study the same effect, see Figure~\ref{fig_chaotic}(b) for the densities considered here.   Another key observation is that certain chaotic sequences have a density which has the same algebraic description as the distribution of known random sequences.
For instance, the 
time evolution of the Logistic map with $r=4$ has the density $\varrho(z)=\left(\pi \sqrt{z(1-z)}\right)^{-1}$ which is the same function as the probability density of the Beta distribution $\mathcal{B}(\alpha/\beta)$ with $\alpha=\beta=0.5$.  In other words, for certain parameter settings chaotic sequences from the Logistic map and random sequences from the Beta distribution are statistically equivalent as they can be described by one and the same distribution. These observations suggest the following experimental setup. We take chaotic and random sequences which have either the same  distribution or different ones.  We test these sequences in otherwise identical PSO implementations. In this way, we can differentiate  the effect that different or same  distributions have on performance for either random or chaotic sequences. 
Our main result obtained by null hypothesis testing with the Wilcoxon rank-sum test is that the underlying distribution is the key factor in performance, while the origin (either chaotic or random) is secondary.

 \section{Experimental setup} \label{sec:exp}
\textit{Particle swarm optimization} (PSO) is a generational population-based algorithm used for calculating optima of single- and multidimensional functions $f(x)$. A PSO particle holds 4 variables for each  generation $t$: its position $x(t) \in \mathbb R^D$ and velocity $v(t)\in \mathbb R^D$ in search space, its own best position (local best position) $p^l(t)$ and the best position of any particle of the population that occurred during all generations (global best position) $p^g(t)$. The particle movement is described by  
\begin{align}
x(t+1) =& x(t) + v(t+1) \label{eq:pos}\\
v(t+1) =&wv(t)+c_1r_1(p^l(t) - x(t))+c_2r_2(p^g(t) - x(t)) \label{eq:vel}.
\end{align}
The parameters  $w$, $c_1$ and $c_2$ are the inertial,  cognitive and social weights. The random variables $r_1,r_2$ are taken from the sequences produced by either the PRNGs or chaotic maps.
At $t=0$, each particle $i$, $i=1,2,\ldots I$,  of  swarm size $I$ is initialized randomly at position $x(0)$ with velocity $v(0)$. The fitness function is evaluated for the position of every particle. The position with the highest fitness that occurred in this evaluation is stored as global best position in each particle.  

In the simulation we take standard values of PSO parameters frequently used: $w = 0.79$, $c_1 = 1.49$, $c_2 = 1.49$, $200$ generations per run, and $1000$ runs per sample. 
The performance of PSO runs is quantified by the mean distance error, which is the mean of the absolute difference in search space between the found best fitness value and the global optimum of the function $f(x)$. 
 A good performance means a low mean distance between the found best result and the actual optimum of the fitness function.

To compare the impact of chaotic and random sequences on PSO the following test functions $f(x)$ from the IEEE CEC 2013 test suite~\cite{li13} are employed:
\begin{itemize}
    \item 1 Equal Maxima (1 dimension)
    \item 2 Uneven Decreasing Maxima (1 dimension)
    \item 3 Himmelblau (2 dimension)
    \item 4 Six-Hump Camel Back (2 dimension)
    \item 5 Shubert (2 dimension)
    \item 6 Vincent (2 dimension)
    \item 7-9 Rastrigin (10, 20 and 30 dimensions)
    \item 10-12 Rosenbrook (10, 20 and 30 dimensions)
    \item 13-15 Sphere (10, 20 and 30 dimensions).
\end{itemize}
\noindent
Random sequences with the following distributions  are used: 
\begin{itemize}
    \item $\mathcal{B}(0.5/0.5)$ Beta distribution
    \item $\mathcal{B}(1/5)$ Beta distribution
    \item $\mathcal{B}(5/1)$ Beta distribution
    \item $\mathcal{N}(0.5/0.1)$ normal distribution. 
\end{itemize} 
\noindent
Chaotic sequences from the following chaotic maps are taken:
\begin{itemize}
    \item Logistic map, $z(k+1)=rz(k)(1-z(k))$, $r=4$
    \item Cubic map, $z(k+1)=rz(k)(1-z(k)^2)$, $r=2.62$
    \item Bellows map, $z(k+1)=rz(k) (1+z(k))^{-6}$, $r=2.0$.
 \end{itemize}
\noindent

\begin{table*}[tb]
  \centering
  \caption{Comparison between chaotic and random sequences for the test function according to Section \ref{sec:test} and the distributions in Section \ref{sec:dist}. Mean and standard deviation of the PSO performance over 1000 runs each. Bold numbers indicate the best performance for each of the 15 test functions. }
    \scalebox{0.7}{
  \begin{tabular}{|c||c|c||c|c||c|c||c|c||c|c||c|c||c|c||}
    \hline
   Test \\ function &  \multicolumn{2}{c}{Logistic Map} & \multicolumn{2}{c}{Cubic Map} & \multicolumn{2}{c}{Bellows Map} & \multicolumn{2}{c}{$\mathcal{B}(0.5/0.5)$} & \multicolumn{2}{c}{$\mathcal{N}(0.5/0.1)$} & \multicolumn{2}{c}{$\mathcal{B}(1/5)$} & \multicolumn{2}{c}{$\mathcal{B}(5/1)$} \\
    \hline
    & $\mu$ & $\sigma$ & $\mu$ & $\sigma$ & $\mu$ & $\sigma$ & $\mu$ & $\sigma$ & $\mu$ & $\sigma$ & $\mu$ & $\sigma$ & $\mu$ & $\sigma$ \\
    \hline
    1 &  0.44e-4 & 0.16e-3 & 0.32e-4 & 0.18e-3 & 0.03e-4 & 0.04e-3 & 0.41e-4 & 0.16e-3 & \textbf{0.00e-4} & 0.00e-3 & 0.02e-4 & 0.03e-3 & \textbf{0.00e-4} & 0.00e-3\\
    \hline
    2 & \textbf{0.10e-2} & 1.05e-2 & 0.26e-2 & 1.95e-2 & 0.15e-2 & 1.38e-2 &  0.16e-2 & 1.48e-2 & 1.12e-2 & 4.12e-2 & 1.13e-2 & 4.24e-2 & 1.19e-2 & 4.24e-2\\
    \hline
    3 &  \textbf{1.58e-1} & 0.69e-2 & \textbf{1.58e-1} & 0.70e-2 & 1.60e-1 & 0.94e-2 & \textbf{1.58e-1} & 0.68e-2 & 1.63e-1 & 1.17e-2 & 1.59e-1 & 0.87e-2 & 1.59e-1 & 0.86e-2\\
    \hline
    4 &  \textbf{1.66e-1} & 2.61e-2 & \textbf{1.66e-1} & 2.59e-2 & \textbf{1.66e-1} & 2.62e-2 & \textbf{1.66e-1} & 2.60e-2 & \textbf{1.66e-1} & 2.61e-2 & 1.68e-1 & 2.54e-2 & \textbf{1.66e-1} & 2.61e-2\\ 
    \hline
    5 & 5.16e-2 & 3.26e-2 & 5.22e-2 & 3.06e-2 & 5.23e-2 & 3.19e-2 &  5.22e-2 & 3.21e-2 &4.93e-2 & 3.09e-2 & \textbf{4.82e-2} & 2.88e-2 & 4.98e-2 & 3.07e-2\\
    \hline
    6 &  3.62e-6 & 0.22e-5 & 3.50e-6 & 0.22e-5 & 4.11e-6 & 1.41e-5 & 3.68e-6 & 0.22e-5 & \textbf{3.44e-6} & 0.22e-5 & 3.49e-6 & 0.22e-5 & 3.45e-6 & 0.22e-5\\
    \hline
    7 &7.74e-2 & 2.11e-2 & \textbf{7.27e-2} & 1.94e-2 & 7.73e-2 & 2.11e-2 &  7.67e-2 & 2.22e-2 & 7.39e-2 & 1.92e-2 & 7.37e-2 & 1.88e-2 & 7.51e-2 & 2.02e-2\\
    \hline
    8 &  8.03e-2 & 1.75e-2 & 7.72e-2 & 1.77e-2 &8.09e-2 & 1.99e-2 & 7.95e-2 & 1.79e-2 & 7.65e-2 & 1.74e-2 & \textbf{7.60e-2} & 1.54e-2 & 7.84e-2 & 1.85e-2\\
    \hline
    9 &  8.38e-2 & 1.78e-2 & 8.02e-2 & 1.73e-2 & 8.47e-2 & 1.98e-2 & 8.35e-2 & 1.77e-2 & 8.02e-2 & 1.84e-2 & \textbf{7.75e-2} & 1.44e-2 & 8.39e-2 & 2.11e-2\\
    \hline
    10 &  7.28e-2 & 1.73e-2 & 7.29e-2 & 1.79e-2 & 8.05e-2 & 1.58e-2 & \textbf{7.20e-2} & 1.80e-2 & 8.93e-2 & 1.28e-2 & 8.87e-2 & 1.21e-2 & 8.65e-2 & 1.35e-2\\
    \hline
    11 & 9.14e-2 & 0.74e-2 & 9.18e-2 & 0.69e-2 & 9.07e-2 & 0.78e-2 & 9.12e-2 & 0.74e-2 &  9.05e-2 & 0.77e-2 & 9.06e-2 & 0.70e-2 & \textbf{9.04e-2} & 0.76e-2\\
    \hline
    12 &  9.06e-2 & 0.63e-2 & 9.05e-2 & 0.66e-2 & 9.10e-2 & 0.66e-2 & 9.12e-2 & 0.61e-2 & \textbf{9.03e-2} & 0.67e-2 & 9.08e-2 & 0.64e-2 & 9.04e-2 & 0.70e-2\\
    \hline
    13 & \textbf{0.01e-4} & 0.02e-4 & 0.03e-4 & 0.05e-4 & 0.05e-4 & 0.38e-4 &  \textbf{0.01e-4} & 0.02e-4 & 0.36e-4 & 1.73e-4 & 0.36e-4 & 1.72e-4 & 1.36e-4 & 2.26e-4\\
    \hline
    14 &  0.64e-2 & 0.28e-2 & \textbf{0.51e-2} & 0.27e-2 & 0.81e-2 & 0.33e-2 & 0.68e-2 & 0.29e-2 & 0.73e-2 & 0.30e-2 & 0.86e-2 & 0.33e-2 & 0.94e-2 & 0.35e-2\\
    \hline
    15 &  1.67e-2 & 0.41e-2 & \textbf{1.55e-2} & 0.40e-2 & 1.83e-2 & 0.43e-2 & 1.71e-2 & 0.42e-2 & 1.67e-2 & 0.43e-2 & 1.91e-2 & 0.44e-2 & 1.97e-2 & 0.44e-2\\
    \hline
  \end{tabular}
  }
    \label{tab:1a}
\end{table*}


\section{Results}
Table \ref{tab:1a} presents a comparison between chaotic and random sequences for the test functions and the distributions in Section \ref{sec:exp}. For each combination of test function and distribution the mean performance and the standard deviation of the performance are given for 1000 PSO runs each. Bold numbers indicate the best performance for each of the 15 test functions. The results vary not much for a given test function; there are several test functions where more than one distribution scores best results, but there  are  also subtle differences in performance. For instance,  PSO driven by sequences associated with the Cubic map performs best for 5 out of the 15 test functions, but Bellows map only scores 1 out of 15  and the Logistic map has 4 out of 15. For the random sequences the results are a little more balanced with PSO driven by sequences from  $\mathcal{B}(0.5/0.5)$ and $\mathcal{N}(0.5/0.1)$ performing best 4 times out of 15, while $\mathcal{B}(1/5)$ and $\mathcal{B}(5/1)$ come up with 3 best results. At least for the selection considered, every distribution is best for at least one test function. Moreover, if we were to compare chaotic sequences from the Cubic map and random sequences from $\mathcal{B}(1/5)$ (or $\mathcal{B}(5/1)$), we might be inclined to conclude that chaos and random give different performances with chaos somewhat being better. In the following, we analyse the performance results and argue that these differences are rather caused by differences in the underlying distribution and to a lesser degree by the fact that the sequences come from either PRNGs or chaotic maps.

\begin{figure}[htb]
\begin{center}
\includegraphics[trim = 15mm 10mm 10mm 8mm,clip, width=8cm,height=6cm]{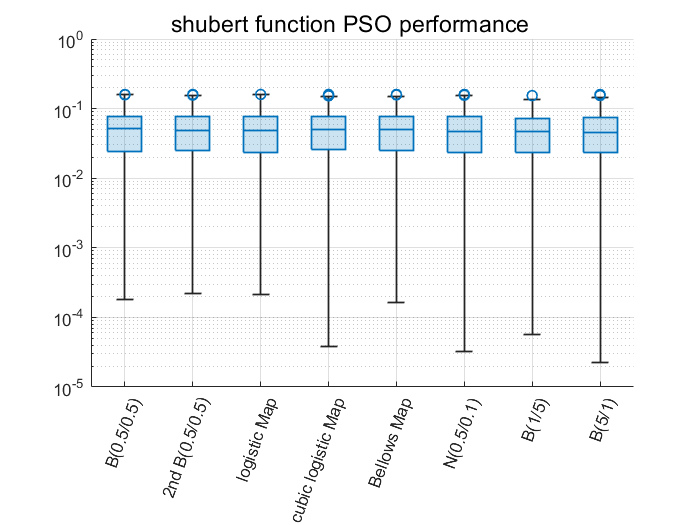}

\caption{Box plot for PSO performance data for the Shubert function.}
\label{fig_shubert}
\end{center}
\end{figure}

A first step is by displaying the distribution of the 1000 PSO runs as boxplots to evaluate not only mean and standard deviation of the results, but also quartiles and outliers. With the two dimensional Shubert function we take an example showing some systematic differences of the PSO performance  which are typical for all test functions considered, see Figure \ref{fig_shubert}. In addition to the data in Table~\ref{tab:1a}, we also give the result of a re-run with $\mathcal{B}(0.5/0.5)$, for which we get the same mean and standard deviation, but a slight difference in the lower whisker. Although the means and standard deviations and also the median, first and third quartiles for all distributions are almost equal,
for PSO run samples that were driven by $\mathcal{B}(0.5/0.5)$, the Logistic map and Bellows map, the lower quartile of the performance data is significantly narrower than their counterparts resulting from  $\mathcal{B}(1/5)$, $\mathcal{B}(5/1)$, $\mathcal{N}(0.5/0.1)$ and the Cubic map. In other words, we may conclude that  PSO performs slightly worse when using $\mathcal{B}(0.5/0.5)$, Logistic map or Bellows map, respectively, as for sequences from the other 3 distributions very good results fall into the quartile group 1. Moreover,  for these 3 distributions the performance is almost the same. Furthermore,  $\mathcal{B}(0.5/0.5)$ in both runs and the Logistic map are most similar.

\begin{table*}[tb]
  \centering
   \caption{Performance comparison between chaotic and random sequences by Wilcoxon rank-sum test for all 15 test function according to Section \ref{sec:test} and the 7 distributions in Section \ref{sec:dist}. For each cell the tuple $[+,-,\thickapprox]$ indicates for $+$ that the sequences from the distribution in the row performs better, for $-$ that the sequences from the distribution in the column performs better, while for $\thickapprox$  there is no difference in performance. Bold numbers indicate comparisons between chaotic and random sequences. }
  \scalebox{0.65}{
  \begin{tabular}{|c||c|c|c|c|c|c||c||}
    \hline
    & $\mathcal{B}(0.5/0.5)$  & \parbox{1cm}{\centering cubic\\map} & \parbox{1cm}{\centering Bellows\\map} & $\mathcal{N}(0.5/0.1)$ & $\mathcal{B}(1/5)$ & $\mathcal{B}(5/1)$ & chaotic vs. random\\
    \hline
    $\mathcal{B}(0.5/0.5)$ & [1,0,14] &   &    &  [4,7,4] &  [5,6,4] &  [5,4,6] &  \\
    \hline
    logistic map & \textbf{[2,1,12]} &    [1,5,9] &  [5,2,8] &  \textbf{[3,6,6]} &  \textbf{[5,6,4]} &  \textbf{[4,5,6]} & 14 ($+$), 18 ($-$), 28 ($\thickapprox$)\\
    \hline
    cubic map & \textbf{[5,2,8]} &   &     [9,1,4] &  \textbf{[5,3,7]} &  \textbf{[5,4,6]} &  \textbf{[7,2,6]} & 22 ($+$), 11 ($-$), 27 ($\thickapprox$)\\
    \hline
    Bellows map & \textbf{[2,4,9]} &   &      &  \textbf{[3,7,5]} &  \textbf{[5,6,4]} &  \textbf{[4,5,6]} & 14 ($+$), 22 ($-$), 24 ($\thickapprox$)\\
    \hline
    $\mathcal{N}(0.5/0.1)$ &  &      &   &  &  [2,3,10] &  [5,3,7] &\\
    \hline
    $\mathcal{B}(1/5)$ &  &   &     &  &  &  [5,2,8] &\\
    \hline \hline
    random vs. chaotic & 7 ($+$), 9 ($-$), 29 ($\thickapprox$)& & & 16 ($+$), 11 ($-$), 18 ($\thickapprox$) & 16 ($+$), 15 ($-$), 14 ($\thickapprox$) & 12 ($+$), 15 ($-$), 18 ($\thickapprox$)& \\ \hline
  \end{tabular}
  }
   \label{tab:wilcoxon}
\end{table*}

Our second analysis step is to compare the performance  with a nonparametric statistical test of the null hypothesis. For this task, the Wilcoxon rank-sum test is widely used to verify performance differences between variants of metaheuristic algorithms. It is here taken to evaluate performance differences between chaotic and random sequences. The two sided test evaluates the null hypothesis that two given data samples are samples from continuous distributions with the same median. We use a 0.05 significance level.
The Wilcoxon rank-sum test is used on every combination of PSO performance samples. Table \ref{tab:wilcoxon} states for each comparison between differently driven PSO the following tuple $[+,-,\thickapprox]$ with:
\begin{itemize}
    \item $+$ \dots number of test functions where the PSO driven by the sequence in the \textit{row} performs better
    \item $-$ \dots number of test functions where the PSO driven by the sequence in the \textit{column} performs better
    \item $\thickapprox$ \dots number of test functions where the  PSO performances are not distinguishable by the Wilcoxon rank-sum test.
\end{itemize}
The results show that performance differences primarily occur when the used sequences differ in the underlying probability distribution. For instance, compare the results for the Logistic map with $\mathcal{B}(0.5/0.5)$, which gives 5 superior, 6 inferior and 4 equal performances, or the results for  $\mathcal{B}(0.5/0.5)$ vs.  $\mathcal{B}(1/5)$ with 4 superior, 5 inferior and 6 equal.
In contrast  the PSO using sequences from  $\mathcal{B}(0.5/0.5)$ and the Logistic map have very small deviations in performance with 1 superior, 2 inferior and 12 equal. This is almost on a par with a re-run with another set of realizations of $\mathcal{B}(0.5/0.5)$, which gives 1 inferior and 14 equal. Another interesting comparison is the overall performance of chaotic sequences vs. random sequences, see the bold numbers in Table \ref{tab:wilcoxon} and the summation in the last row and column. Here, we find for chaos vs. random that the Cubic map gives 22 superior, 11 inferior and 27 equal performances, while the Logistic and Bellows map score 14 superior vs. 18 (and 22) inferior and 28 (and 24) equal performances. Similar results  can be recorded from the perspective of random sequences vs. chaotic sequences. For instance,  $\mathcal{B}(0.5/0.5)$ has 7 superior, 9 inferior and 29 equal performances, but  $\mathcal{N}(0.5/0.1)$ scores 16 superior, 11 inferior and 18 equal. To summarize some chaotic sequences perform better than random, but there are also some random sequences which perform better than chaos, which agrees with previously reported result~\cite{ala09,chen18,gan13,tian19,plu14,zel22,kuang14,rong10,yang12}. In view of the results presented here, however, it does not appear plausible to assume the presence of general and systematic differences in performance between chaotic and random sequences irrespective and  independent of taking into account the underlying distribution. Given the fact that for random and chaotic sequences with the same distribution, Logistic map and $\mathcal{B}(0.5/0.5)$, we get the smallest difference in performance, while for all other combinations of different sources, we get different performances, the results rather support the conclusion that the underlying distribution rather than the origin is the main influential factor   in PSO performance.

\section{Discussion}
 There is an ongoing debate about using random and chaotic sequences for driving population-based metaheuristic search algorithms and the effect on behaviour and performance such a selection has~\cite{capo03,chen18,gag21,xu18,ma19,gan13,tian19,plu14,zel22}. This paper adds to this discussion with the focus on PSO and proposed the following approach. We not only use random sequences associated with different  probability distributions, but also compare to chaotic sequences with different  density distributions and complete the experimental setup by random and chaotic sequences with the same distribution. This enables us to differentiate between differences in the origin of the sequences (PRNGs or chaotic maps) and statistical differences expressed by the underlying distributions. 
 
  These differences in statistical properties can potentially have profound influences on PSO  search dynamics, and thus on behaviour and performance. The particle movement directly depends on the sequences as successive particle positions are calculated via cognitive and social components  weighted by the elements from the sequences. The distribution associated with the sequences defines global, long-term search preferences of the particle swarm. If, for instance, the distribution is rather bell shaped, as a normal distribution, 
then the search is more likely centered to the interior of the search space region where the swarm is currently situated.
If, on the other hand, the distribution is rather U-shaped, as Beta distributions with $\alpha=\beta<1$, the search more likely concentrates on the exterior of the current region.  How such a varying search bias interacts with unimodal or multimodal  functions of the optimization problem causes different types of swarm diversity decline. By this mechanism does the
distribution of the numbers in the random and chaotic sequences influence the ratio of exploration and exploitation during the PSO run and thus behavior and performance. 

In the experimental setup, we considered 3 sources of chaotic sequences, Logistic, Cubic and Bellows map, and 4 sources of random sequences, a normal distribution
$\mathcal{N}(0.5/0.1)$ and 3 Beta distributions, $\mathcal{B}(0.5/0.5)$, $\mathcal{B}(1/5)$ and $\mathcal{B}(5/1)$. With this selection, 
we took distributions which are not constant but have shape over the sample space. This is in contrast to an uniform distribution, which is constant and produces random sequences which are equally spread on a fixed interval. For such a type of random distribution results similar to those discussed here are known.
The invariant density of the tent map equals an uniform distribution and    
differences in performance between chaotic sequences generated by the tent map as compared to sequences generated by the Logistic map have been shown~\cite{kuang14,rong10}, which is consistent with and supplements our results.

\section{Conclusions}
We considered PSO and address the question whether or not using chaotic or random sequences really causes different performance. Thus, our discussion is not so much about what sequence performs best, but rather about identifying reasons why  some sequences are better than others.
Our findings cast considerable doubt on previous results which compared random to chaos and suggested that there are intrinsic differences in PSO performance,   irrespective and independent of the associated distributions. We may choose either random or chaotic sequences to drive PSO and observe differences in performance. But most likely these differences occur because the underlying distributions of the chosen  sequences are different.  In other words, if chaotic sequences for a certain optimization problem (or also for a certain collection  of optimization problems) lead to  better performance then this is primary the case because the distribution is more fitting to the problem.


\begin{thebibliography}{1}

\bibitem{ala09} Bilal Alatas, Erhan Akin, and A. Bedri  Ozer. 2009. Chaos embedded particle swarm optimization algorithms. Chaos, Solitons \& Fractals, 40(4), 1715-1734. 

\bibitem{chen18}  Ke Chen,  Fengyu Zhou, and Aling Liu. 2018. Chaotic dynamic weight particle swarm optimization for numerical function optimization. Knowledge-Based Systems, 139, 23-40.

\bibitem{capo03} Riccardo Caponetto,  Luigi Fortuna,  Stefano Fazzino, and
Maria Gabriella Xibilia. 2003. Chaotic sequences to improve the performance of evolutionary algorithms. IEEE Trans. Evolut. Comp. 7, 289-304.


 
\bibitem{gag21} Iannick Gagnon, Alain April, and Alain  Abran. 2021. An investigation of the effects of chaotic maps on the performance of metaheuristics. Engineering Reports, 3(8), e12369.

\bibitem{gan13} Amir Hossein Gandomi, Gun Jin Yun, Xin-She Yang, and Siamak Talatahari. 2013. Chaos-enhanced accelerated particle swarm optimization. Communications in Nonlinear Science and Numerical Simulation, 18(2), 327-340.



\bibitem{kuang14} Fangjun Kuang,  Zhong Jin, Weihong Xu, and Siyang Zhang. 2014.  A novel chaotic artificial bee colony algorithm based on tent map. In: Proc. 2014 IEEE Congress on Evolutionary Computation (CEC),  IEEE, Pistacaway,  235--241.

\bibitem{li13} Xiaodong Li, Andries Engelbrecht, and Michael G. Epitropakis. 2013. Benchmark functions for CEC’2013 special session and competition on niching methods for multimodal function optimization. RMIT University, Evolutionary Computation and Machine Learning Group, Australia, Tech. Rep.

\bibitem{liu05} Bo Liu, Ling Wang, Yi-Hui Jin, Fang Tang,  and De-Xian Huang. 2005. Improved particle swarm optimization combined with chaos. Chaos, Solitons \& Fractals, 25(5), 1261-1271.



\bibitem{ma19} Zhiteng Ma, Xianfeng Yuan, Sen Han, Deyu Sun,
and Yan Ma. 2019. Improved chaotic particle swarm optimization algorithm with more symmetric distribution for numerical function optimization. Symmetry, 11(7), 876.




\bibitem{plu14} Michal Pluhacek, Roman Senkerik, and Ivan  Zelinka. 2014. Particle swarm optimization algorithm driven by multichaotic number generator. Soft Comput. 18(4), 631--639. 





\bibitem{rong10} Hua Rong. 2010. Study of adaptive chaos embedded particle swarm optimization algorithm based on skew tent map. In Proc. 2010 International Conference on Intelligent Control and Information Processing,  IEEE, Piscataway, 316-321.



\bibitem{tian19} Dongping Tian, Xiaofei Zhao, and Zhongzhi Shi. 2019.
Chaotic particle swarm optimization with sigmoid-based acceleration coefficients for numerical function optimization.
Swarm and Evolutionary Computation,
51,
100573.




\bibitem{xu18} Xiaolong Xu, Hanzhong Rong, Marcello Trovati, Mark Liptrott, and  Nik Bessis. 2018. CS-PSO: chaotic particle swarm optimization algorithm for solving combinatorial optimization problems. Soft Comput. 22(3), 783-795.

\bibitem{yang12} Cheng-Hong Yang, Sheng-Wei Tsai, Li-Yeh Chuang, and Cheng-Huei Yang. 2012. An improved particle swarm optimization with double-bottom chaotic maps for numerical optimization. Applied Mathematics and Computation, 219(1), 260-279.


\bibitem{zel22} Ivan Zelinka, Quoc Bao Diep, Vaclav Snasel, Swagatam Das, Giacomo Innocenti, Alberto Tesi, Fabio Schoen, and Nikolai V. Kuznetsov. 2022. Impact of chaotic dynamics on the performance of metaheuristic optimization algorithms: An experimental analysis. Information Sciences, 587, 692-719.



























\end{thebibliography}
\end{document}